\pdfoutput=1 
\documentclass[a4paper, 12pt]{article}
\usepackage{jheppub}
\usepackage{preamble}
\begin{document}
\title{The SUSY Index Beyond the Cardy Limit}
\author{Ohad Mamroud}
\affiliation{Department of Particle Physics and Astrophysics, \\[.0em]
Weizmann Institute of Science, Rehovot 7610001, Israel}
\emailAdd{ohad.mamroud@weizmann.ac.il}
\abstract{We analyze a set of contributions to the superconformal index of 4d $\cN = 4$ $SU(N)$ super Yang-Mills using the Bethe Ansatz approach. These contributions dominate at the large $N$ limit, where their leading order in $N$ reproduces various supersymmetric Euclidean black hole saddles in the dual theory, and they also dominate for finite $N$ in high temperature Cardy-like limits. We compute the $O(N^0)$ terms, including those exponentially suppressed in the Cardy limit, and show that there are no $1/N$ corrections beyond them. Under certain assumptions, it implies that the gravitational perturbative series around these black hole saddles is 1-loop exact.}

\maketitle

\section{Introduction}
The last few years have seen immense progress in the ability to compute the entropy of black holes and quantum gravity partition functions through their holographic dual conformal field theories, at least in supersymmetric cases. The pioneering works of \cite{Benini:2015eyy,Benini:2018ywd,Cabo-Bizet:2018ehj,Choi:2018hmj} had succeeded in reproducing the leading contribution to the black hole partition function and reproduced the black hole entropy, while subsequent works had matched both perturbative and non-perturbative effects, in various dimensions and settings \cite{Honda:2019cio, ArabiArdehali:2019tdm, Kim:2019yrz, Cabo-Bizet:2019osg, Amariti:2019mgp, Lezcano:2019pae, Lanir:2019abx, Cabo-Bizet:2019eaf, ArabiArdehali:2019orz, Cabo-Bizet:2020nkr, Murthy:2020rbd, Agarwal:2020zwm, Benini:2020gjh, GonzalezLezcano:2020yeb, Copetti:2020dil, Cabo-Bizet:2020ewf, Amariti:2020jyx, Hosseini:2016tor, Benini:2016hjo, Benini:2016rke, Hosseini:2016cyf, Cabo-Bizet:2017jsl, Azzurli:2017kxo, Hosseini:2017fjo, Benini:2017oxt, Hosseini:2018uzp, Crichigno:2018adf, Hosseini:2018usu, Suh:2018szn, Fluder:2019szh, Gang:2019uay, Kantor:2019lfo, Choi:2019zpz, Bobev:2019zmz, Nian:2019pxj, Benini:2019dyp, David:2021qaa, David:2020ems, Amariti:2022nvn, Hosseini:2021mnn, Hong:2021bzg, Goldstein:2020yvj, Jejjala:2021hlt, Jejjala:2022lrm}.

In this work, we will concentrate on the case of $\cN = 4$ $SU(N)$ super Yang-Mills theory. With a specific choice of chemical potentials, the partition function of this theory essentially becomes a protected quantity, the superconformal index, which is sensitive to 1/16-BPS states in the theory. On the other hand, the gravitational partition function of the dual theory can be computed in a semi-classical manner as a sum over the Euclidean saddles of type IIB supergravity on asymptotically $\AdS_5 \times S^5$ with matching boundary conditions.

Previous works \cite{Benini:2018ywd,Cabo-Bizet:2018ehj,Choi:2019zpz,Choi:2021rxi} had shown various ways in which information about different Euclidean gravitational saddles, describing black hole and orbifold geometries, can be extracted from the superconformal index - either by showing that at large $N$ specific contributions to the index scale as these geometries, or by taking various ``high temperature'' Cardy-like limits where on the gravity side one of these geometries becomes dominant, and calculating the asymptotic behavior of the index in that limit.

In the present paper we concentrate on the field theory side, and compute a set of discrete contributions\footnote{The very same ones that after taking the aforementioned limits would have a gravitational interpretation.} to the index using the Bethe Ansatz approach, at finite $N$ and $\tau$, where $\tau$ denotes one of the chemical potentials, to be defined below. We then show that in the large $N$ limit the perturbative contribution of our solution in $1/N$ truncates at $O(N^0)$, and explicitly compute these terms. Moreover, we compute the contribution at finite $N$ and in the Cardy limit $\tau\to 0$, including $O(e^{-1/\tau})$ terms up to $O(e^{-N/\tau})$ and find that it takes a particularly compact form. Since the leading order contributions we compute match those of black hole saddles on the gravity side, we conjecture that the gravitational perturbation theory around these saddles truncates as well.

Given some supercharge $\cQ$, the superconformal index of the theory is
\begin{equation}
\cI = \Tr\left[(-1)^F e^{-\beta\{\cQ,\bar \cQ\} + 2\pi i \tau \left(J_1 + \frac{1}{2}R_3 \right)  + 2\pi i \sigma \left(J_2 + \frac{1}{2}R_3\right) + 2\pi i \Delta_1 \fq_1 + 2\pi i \Delta_2\fq_2} \right] \;,
\end{equation}
where $J_{1,2} + \frac{1}{2} R_3$ and $\fq_{1,2}$ are conserved charges that commute with $\cQ$, for further details see Appendix \ref{app:Index def}. The index can be expressed as an integral over the holonomies $\{u_i\}$ of the $SU(N)$ gauge field around the thermal cycle. As reviewed in Section \ref{sec:Bethe Ansatz Review} and following \cite{Benini:2018mlo}, when $\tau = \sigma$ and after some manipulation this integral can be evaluated using the residue theorem. The residues are solutions to some complicated equations, \eqref{eq:reduced_BAE}, termed the Bethe Ansatz equations. Eventually, the index can be expressed as a sum over configurations of the holonomies, schematically
\begin{equation}
\cI(\tau, \Delta_1, \Delta_2) = \kappa \sum_{\hat u \,\in\, \text{BAEs}} \cZ(\hat u; \dlt, \tau) \, H(\hat u; \dlt, \tau)^{-1} + \text{vanishing Jacobian}\;,
\end{equation}
where $\cZ$ and $\kappa$ come from the integrand and $H$ is the Jacobian relating the integration variable to the Bethe Ansatz equations. Some residues might have a vanishing Jacobian, and their contribution to the index will be described in an upcoming paper \cite{Aharony:2023}.

A general classification of the solutions is still beyond reach, but there is a known family of solutions called the Hong-Liu solutions \cite{Hong:2018viz}. They are given by symmetric configurations of the $N$ holonomies $u_i$ on a lattice of modular parameter $\tau$
\begin{equation}
\label{eq:Hong Liu solutions}
u_{j} = u_{\hat{\jmath}\hat{k}} = \bar{u}+\frac{\hat{\jmath}}{m}+\frac{\hat{k}}{n}\left(\tau+\frac{r}{m}\right)\ ,\qquad \hat\jmath = 0,\dots,m-1\;,\qquad \hat k = 0,\dots,n-1 \;,
\end{equation}
where $N = m\cdot n$, $r = 0,\dots,n-1$ and $\bar u$ is chosen such that $\sum_j u_j = 0$, as the holonomies are of an $SU(N)$ matrix. These configurations are sometimes denoted by the triplet $\{m,n,r\}$. Their contribution to the index is denoted by $\cI_{\{m,n,r\}}$, and there could be additional contributions as well,
\begin{equation}
    \cI = \sum_{\{m,n,r\}} \cI_{\{m,n,r\}} + \text{others} \;.
\end{equation}

The partition function of the theory on $S^1 \times S^3$ with a particular choice of chemical potentials is proportional to its superconformal index \cite{Closset:2013vra}. Since the partition function should match that of the dual gravitational theory, \cite{Aharony:2021zkr} argued\footnote{Following \cite{Benini:2018ywd} who first used the method to match one of the gravitational saddles and reproduced the entropy of the Lorentzian BPS black hole, and \cite{Cabo-Bizet:2018ehj,Choi:2018hmj} who achieved this by different methods. See also \cite{GonzalezLezcano:2020yeb} for the expansion to this order of the BA solution.} that in the large $N$ limit each Hong-Liu configuration with $n \sim N$ and $m,r \ll n$ reproduces the action of a different Euclidean geometry, in the sense that the ``action'' of an HL configuration is
\begin{equation}
\log \cI_{\{m,n,r\}} = -c_{\{m,n,r\}} N^2 + \log N + O(N^0) \;,
\end{equation}
where $c_{\{m,n,r\}} N^2$ is the on-shell action of the corresponding gravitational saddle. These geometries describe (complex) Euclidean black holes and their orbifolds. The argument hinges on evaluating $\cZ$ and $\kappa$ in the large $N$ limit, and bounding the contribution of the Jacobian $H$. Moreover, $\cZ$ contains $O(e^{-c_{D_3}N})$ terms, corresponding to the on-shell action of additional saddles containing arbitrary number of wrapped supersymmetric D3 branes on top of these geometries.

In this paper we exactly compute the Jacobian $H$ for the Hong-Liu solutions. When $n=N\to\infty, r \ll  n, m = 1$, which is the large $N$ limit corresponding to the various black hole saddles without orbifolds, we find that the series truncates\footnote{This was conjectured in \cite{Aharony:2021zkr}, but was not proven there.} at order $N^0$ up to exponential corrections
\begin{equation}
\log \cI_{\{1,N,r\}} \xrightarrow{N\to\infty} -c_{\{1,N,r\}} N^2 + \log N + c^{(0)}_{\{1,N,r\}} + O(e^{-N c_{D3}}) \;.
\end{equation}
We conjecture that this result carries over to the gravity side, meaning that the gravitational perturbative series truncates at order $O(G_N^0)$ and is one-loop exact around these saddles. Moreover, the new exponentially small corrections we have found are of the same order of magnitude as the exponential correction found before, and therefore can again be considered as arising from saddles with additional branes. 

It should be stressed that what we actually compute in this paper is the contribution of a set of Bethe Ansatz solutions to the index, whose leading order behavior matches that of the gravitational saddles. We have not performed any new gravitational computation. However, these are the leading contributions to the index in various Cardy-like limits (see \cite{Cassani:2021fyv,ArabiArdehali:2021nsx}). As long as there is no other BA contribution whose action differs from $\log \cI_{\{1,N,r\}}$ by $O(N^0)$, our conjecture should hold. Amongst the Hong-Liu family this is indeed the case. It seems likely that this is the case in general - any other BA configuration would differ by at least one of the eigenvalues $u_i$, and thereby change $4N$ out of the $4N^2$ terms in the relevant $\log \cZ$, generically changing it by $O(N)$.

We are also able to analyze the theory in the Cardy limit $\tau \to 0$, in which the $\{1,N,0\}$ HL configuration dominates. Up to order $O(e^{-1/\tau})$ our expression agrees with those of \cite{GonzalezLezcano:2020yeb,ArabiArdehali:2021nsx,Cassani:2021fyv,Cabo-Bizet:2019osg,Kim:2019yrz}. The only corrections up to $O(e^{-N/\tau})$ come from the Jacobian. The final expression takes the rather compact form\footnote{Again, we assume that the $O(e^{-1/\tau})$ corrections come from the leading Bethe Ansatz configuration and not from another, unknown, contribution.}
\begin{multline}
\label{eq:Cardy_limit_res}
\cI \xrightarrow{\tau \to 0}
\frac{e^{\frac{\pi i}{12}}}{\tau \cI_{U(1)}} N e^{-\frac{\pi i N^2 [\dlt_1]_\tau [\dlt_2]_\tau [\dlt_3]_\tau}{\tau^2}} \prod_{k=1}^\infty \left[\frac{1-\tilde{q}^k}{\left(1 - \tilde{y}_1^k\right)\left(1 - \tilde{y}_2^k\right)\left(1 - \tilde{y}_3^k\right)}\right]^2 \\ \times \left(1 + O\left(\tilde{q}^N,\tilde{y_a}^N\right)\right)\;.
\end{multline}
where $\tilde{q} = e^{-2\pi i/\tau}$ and $\tilde{y}_a = e^{2\pi i [\dlt_a]_\tau/\tau}$. Exact details and definition of $[\Delta]_\tau$ are given in Section~\ref{sec:Index in limits}. The final 1-loop determinant looks vaguely similar\footnote{We thank Francesco Benini for pointing our attention to the similarity.} to the multi-graviton partition function of \cite{Kinney:2005ej}, $\cI_\infty = \prod_{k=1}^\infty \frac{(1-q^k)^2}{(1-y_1^k)(1-y_2^k)(1-y_3^k)}$, where $y_a = e^{2\pi i \Delta_1}, q = e^{2\pi i \tau}$, but with the modular transformed fugacities and with a different power of the denominator. It would be interesting to understand if there's a deeper reason for the similarity.

These results raise several questions. The first is, of course, can they be verified by any of the other approaches used to analyze the superconformal index? A direct Cardy limit analysis seems quite hard. One would need to re-sum the exponentially suppressed terms to find the form \eqref{eq:Cardy_limit_res}. However, computing the first few exponentially suppressed terms in the Cardy limit would test our assumptions about the contribution of any other BA solutions. Another method is the large $N$ saddle point analysis of the $\cN = 4$ index matrix model, initiated in \cite{Choi:2021rxi}, in which one of the saddles reproduces the exponential term in \eqref{eq:Cardy_limit_res} for the equal angular velocities case. Our results imply that the perturbative expansion around this saddle, combined with terms of lower order in $N$ in the potential used in \cite{Choi:2021rxi}, should also truncate.

A second, more ambitious question would be whether there is an argument from the gravity side for this apparent one loop exactness in $G_N$, maybe through localization on the supergravity side \cite{Dabholkar:2014wpa,Hristov:2018lod,Hristov:2019xku,Hristov:2021zai,Iliesiu:2022kny}. Lastly, we note that a truncation of similar yet different spirit was found numerically in \cite{Bobev:2022eus,Bobev:2022jte,Bobev:2022wem} for the topologically twisted index of the three dimensional ABJM theory and for its superconformal index in the Cardy limit. It would be nice to understand whether this is a generic feature of supersymmetric partition functions, or merely a coincidence.

The paper is organized as follows. Section \ref{sec:Bethe Ansatz Review} contains a brief overview of the Bethe Ansatz approach. Section \ref{sec:Index in limits} describes the explicit behavior of the Hong-Liu solutions in the large $N$ and Cardy limits, up to terms exponentially suppressed in $N$. Appendix \ref{app:Index def} includes definitions of the index and our conventions. Appendix \ref{app:simplifying HL contribution} contains the explicit computation of the Hong-Liu contributions. Appendix \ref{app:Special functions} contains definitions of the special functions we use throughout the paper.

\section{The Bethe ansatz approach}
\label{sec:Bethe Ansatz Review}
Here we review the computation of the superconformal index for $\cN = 4$ super Yang-Mills (SYM) using the so-called Bethe Ansatz approach. The theory is an $SU(N)$ superconformal gauge theory, with an $SU(4)_R$ R-symmetry, and when put on the cylinder $\bR \times S^3$ it has an additional $SO(4)$ symmetry due to the isometries of the underlying manifold. It has six real adjoint scalars in the $\mathbf{6}$ of the R-symmetry group and four adjoint Weyl fermions in its $\mathbf{4}$. Turning on chemical potentials for two elements of the Cartan which commute with a specific supercharge, and equal chemical potentials for the two angular momenta (see Appendix \ref{app:Index def} for exact details) the superconformal index can be written in the integral form
\begin{equation}
\label{eq:Index integral formula}
    \cI = \frac{\cI_{U(1)}^{N-1}}{N!} \oint_{\bT^{N-1}}  \left(\prod_{i=1}^{N-1}\frac{dz_i}{2\pi i z_i}\right) \cZ\left(\{u_i\};\dlt,\tau\right) \;,
\end{equation}
where $\bT^{N-1}$ is the torus $\{\forall i: |z_i| = 1\}$, $z_i = e^{2\pi i u_i}$, $u_{ij} = u_i - u_j$, and $u_N = -\sum_{i=1}^{N-1}u_i$ so the $u$'s can be thought of the eigenvalues of an $N\times N$ traceless matrix. We also use the fugacity $q = e^{2\pi i \tau}$. The integrand and prefactor are
\begin{equation}
\label{eq:def integrand}
    \cZ\left(\{u_i\};\dlt, \tau\right) = \frac{\prod_{a=1}^3 \prod_{i\neq j}^N \wt\Gamma(\dlt_a + u_{ij};\tau,\tau)}{\prod_{i\neq j}^N \wt\Gamma(u_{ij};\tau,\tau)} \;, \qquad \cI_{U(1)} = (q;q)_\infty^2\prod_{a=1}^3 \wt\Gamma\left(\dlt_a;\tau,\tau\right) \;.
\end{equation}
The chemical potentials satisfy $2\tau - \sum_{a=1}^3 \dlt_a \in \bZ$. The function $\wt\Gamma$ is the elliptic Gamma function and $(q;q)_\infty$ is the Pochhammer symbol, see Appendix \ref{app:Special functions}. 

The Bethe Ansatz approach uses the quasi-periodicity of $\wt\Gamma$, \eqref{eq:elliptic gamma quasi periodicity}. There is an elliptic function\footnote{The explicit form of this function is 
\begin{equation}
    \hat Q_i = \frac{Q_i}{Q_N} \;, \quad
    Q_i = e^{2\pi i \sum_{j=1}^N u_{ji}}\prod_{j\neq i}^N \prod_{a=1}^3  \frac{\theta_0(\dlt_a + u_{ji};\tau)}{\theta_0(\dlt_a - u_{ji};\tau)} \;.
\end{equation}
Not all the $Q_i$'s are independent, $\prod_{i=1}^{N} Q_i = 1$.} $\hat Q_i\left(\{u_i\};\dlt,\tau\right)$ with periods $1,\tau$ for which 
\begin{equation}
\label{eq:shifting Z}
    \hat Q_i\left(\{u_i\};\dlt,\tau\right) \cZ\left(\{u_i\};\dlt, \tau\right) = \cZ\left(\{\tilde u_i\};\dlt, \tau\right) \;,
\end{equation}
where $\tilde u = (u_1,\dots,u_{i-1},u_i-\tau,u_{i+1},\dots,u_{N-1},u_N+\tau)$. Now comes the trick. Multiply both the numerator and the denominator of \eqref{eq:Index integral formula} by the same factor $\prod_{i=1}^{N-1} (1 - \hat Q_i)$ to get
\begin{equation}
\begin{aligned}
    \cI &= \frac{\cI_{U(1)}^{N-1}}{N!} \oint_{\bT^{N-1}}  \left(\prod_{i=1}^{N-1}\frac{dz_i}{2\pi i z_i}\right) \frac{\prod_{i=1}^{N-1} (1-\hat Q_i)}{\prod_{i=1}^{N-1} (1-\hat Q_i)} \cZ\left(\{u_i\};\dlt, \tau\right) \;, \\
    &= \frac{\cI_{U(1)}^{N-1}}{N!} \oint_{\cC}  \left(\prod_{i=1}^{N-1}\frac{dz_i}{2\pi i z_i}\right) \frac{\cZ\left(\{u_i\};\dlt, \tau\right))}{\prod_{i=1}^{N-1} (1-\hat Q_i)} \;,
\end{aligned}
\end{equation}
where in the last line we've used the property \eqref{eq:shifting Z} to change the contour integral into the new contour $\cC = \bigcup_{i=1}^{N-1} \left(\{|z_i| = 1\} - \{|z_i| = q^{-1}\} \right)$, where the sign denotes the direction of the contour. Evaluating the integral by the residue theorem, one finds that poles of the numerator are canceled by poles of the denominator, such that the only poles of the integrand come from zeroes of the denominator. A more thorough analysis will appear in an upcoming paper \cite{Aharony:2023}, but a simple sufficient condition is the vanishing of all the terms of the denominator
\begin{equation}
    \hat Q_i = 1 \;, \qquad \forall i = 1, \dots, N-1 \;.
\end{equation}
These are called\footnote{Another interpretation for these equations was initially given by \cite{Closset:2017bse}.} the \emph{Bethe Ansatz equations}. If the solutions are isolated first order zeros of $1 - \hat Q_i$ and the Jacobian for the change of variables from $\{z_i\}$ to $\{\hat{Q}_i\}$ is well defined, then the contribution to the integral is the inverse of the Jacobian times the residue at the pole (which is simply the numerator evaluated at the pole). Finally, the Bethe Ansatz approach to evaluating the index takes the form
\begin{equation}
\label{eq:BA formula}
\cI(q, y_1, y_2) = \frac{\cI_{U(1)}^{N-1}}{N!} \sum_{\hat u \,\in\, \text{BAEs}} \cZ(\hat u; \dlt, \tau) \, H(\hat u; \dlt, \tau)^{-1} + \text{vanishing Jacobian}\;,
\end{equation}
where the sum is over well-behaved solutions to the Bethe Ansatz equations, and the last term denotes contributions where the Jacobian vanishes. Such contributions will be discussed in \cite{Aharony:2023}. Here $\cZ$ and $\cI_{U(1)}$ are defined as in \eqref{eq:def integrand}, and $H$ is the Jacobian,
\begin{equation}
\label{eq:Jacobian def}
    H = \det\left[ \frac1{2\pi i} \, \parfrac{(Q_i / Q_N)}{u_j} \right]_{i,j=1, \dots, N-1} \,\equiv\, \det(\cA_{ij}) \;, 
\end{equation}
we note that $\cI_{U(1)}$ is the index of the free $U(1)$ $\cN = 4$ theory.

In \cite{Benini:2021ano} it was argued that the contribution of many solutions to the Bethe Ansatz equations cancel each other, such that effectively only the solutions to the \emph{reduced Bethe Ansatz equations}
\begin{equation}
\label{eq:reduced_BAE}
    Q_i = (-1)^{N-1} \;,\qquad i = 1,\dots,N-1 
\end{equation}
contribute to the index. There is a known family of solutions to these equations \cite{Hong:2018viz}. In order to find them, note that $Q_i$ is invariant to shifting any of the $u$'s by $1$ or by $\tau$, meaning that the $u$'s naturally live on a torus. The equations are then automatically solved if one distributes the $u$'s symmetrically on the torus, meaning that for every $u_i$ and $u_j$, there exists a $u_k$ such that $u_{ij} = -u_{ik}$ on the torus.\footnote{For odd $N$ the factors in $Q_i$ associated with $u_{ij}$ and $u_{ik}$ then automatically cancel each other, and the proof is complete. For even $N$, a symmetric configuration still leaves for each $u_i$ one $u_j$ for which $u_{ij} = u_{ji}$ on the torus, which are exactly the half periods of the torus. Their contribution to $Q_i$ is exactly $-1$, so it also solves the reduced Bethe Ansatz equations.} These solutions are termed the \emph{Hong-Liu solutions}, whose explicit form is \eqref{eq:Hong Liu solutions}.

\section{The Cardy and large $N$ limits}
\label{sec:Index in limits}
The contribution of the $\{m,n,r\}$ Hong-Liu solution to the index is computed in Appendix \ref{app:simplifying HL contribution}, and turns out to be\footnote{We use a notation where notation $\ck\tau = m\tau + r$ for the $\{m,n,r\}$ solution, and we will use two different conventions for the chemical potentials, $\bdlt_{1,2} = \dlt_{1,2}$ and $\bdlt_3 = \bdlt_1 + \bdlt_2$ or $\dlt_3 = 2\tau - \dlt_1 - \dlt_2 - 1$, such that $\dlt_3 = 2\tau - \bdlt_3 - 1$. Some computations are easier using the barred notation and some using the unbarred one. Only the unbarred notation is used in \cite{Aharony:2021zkr}.}
\begin{equation}
\label{eq:cI Hong Liu Contribution}
\begin{aligned}
    \log \cI_{\{m,n,r\}} &= \sum_{a=1}^{3}\eta_{a} \left[m\log\widetilde{\Gamma}\left(m\bdlt_{a};\frac{\check{\tau}}{n},\frac{\check{\tau}}{n}\right) + (m-N)\log\theta_{0}\left(m\bdlt_{a};\frac{\check{\tau}}{n}\right)\right] \\ &\qquad + N\log\left(\check{q}^{\frac{1}{n}},\check{q}^{\frac{1}{n}}\right)_{\infty}^{2} - \log \cI_{U(1)} + \log N - N\log n \\
    &\qquad - \left(N-1\right)\log b_{N} - \log\det\left(1+\frac{b_{N}^{-1}}{N}\cB^{-1}\cC\right) \;,
\end{aligned}
\end{equation}
where $\eta_a = (1,1,-1)$, $\ck q = e^{2\pi i \ck\tau}$, the special functions are defined in Appendix \ref{app:Special functions}, and
\begin{equation}
\begin{aligned}
    b_N &= -4 + \frac{1}{\ck\tau}\sum_{a=1}^3\left[2m\dlt_a + 1 + \cG\left(0,\frac{N\dlt_a}{\ck\tau};-\frac{n}{\ck\tau}\right)\right] \;,
    \\
    \cB_{ij} &= \delta_{ij}+1 \;, \qquad \cB^{-1}_{ij} = \delta_{ij} - \frac{1}{N} \;, \\
    \cC_{ij} &= \sum_{a=1}^{3}\left[\cG\left(u_{iN},\dlt_{a};\tau\right) + \cG\left(u_{jN}, \dlt_{a};\tau\right) - \cG\left(u_{ij},\dlt_{a};\tau\right) - \cG\left(0,\dlt_{a};\tau\right)\right] \;.
\end{aligned}
\end{equation}
We compute $\log\det\left(1+\frac{b_{N}^{-1}}{N}\cB^{-1}\cC\right)$ by diagonalizing the matrix in Appendix \ref{app:simplifying HL contribution}. The explicit result is given for the cases $m=1$, where we find that the eigenvalues of the matrix $\cB^{-1} \cC$ are practically the discrete Fourier transform of $\cG(\cdot,\frac{\Delta}{\ck\tau};-\frac{1}{\ck\tau})$, which is closely related to the logarithmic derivative of $\theta_0$. We comment on the $m \ll n$ case in the appendix.

\paragraph{Large $n$}
Let us look at the $\{m,n,r\}$ configuration when $n \sim O(N) \to \infty$ and $m, r \ll n$. Then using the modular identities for the various special functions
\begin{equation}
\label{eq:modular transforms special functions}
\begin{aligned}    \log\theta_{0}\left(m\bdlt_{a};\frac{\check{\tau}}{n}\right) &= \log\theta_{0}\left(\frac{N\bdlt_{a}}{\check{\tau}};-\frac{n}{\check{\tau}}\right) + \frac{\pi i}{2} + \pi im\bdlt_{a} - \frac{\pi i\check{\tau}}{6n} \\
    &\qquad\qquad\qquad\qquad\qquad - \frac{n\pi i}{\check{\tau}}\left(\left(m\bdlt_{a}\right)^{2} + \left(m\bdlt_{a}\right)+\frac{1}{6}\right)  \;,\\
    \log\widetilde{\Gamma}\left(m\bdlt_{a};\frac{\check{\tau}}{n},\frac{\check{\tau}}{n}\right) &= -\pi i\mathcal{Q}\left(m\bdlt_{a};\frac{\check{\tau}}{n},\frac{\check{\tau}}{n}\right) - \log\theta_{0}\left(\frac{N\bdlt_{a}}{\check{\tau}};-\frac{n}{\check{\tau}}\right) \\
    & \qquad\qquad\qquad\qquad\qquad\qquad\qquad  + \sum_{k=0}^{\infty}\log\frac{\psi\left(\frac{k+1+m\bdlt_{a}}{\check{\tau}}n\right)}{\psi\left(\frac{k-m\bdlt_{a}}{\check{\tau}}n\right)} \;,
    \\
    \log\left(\check{q}^{\frac{1}{n}},\check{q}^{\frac{1}{n}}\right)_{\infty}^{2} &= \frac{\pi i}{2} - \log\frac{\check{\tau}}{n} - \frac{\pi i}{6}\left(\frac{\check{\tau}}{n} + \frac{n}{\check{\tau}}\right) + \log\left(\tilde{q}^{n};\tilde{q}^{n}\right)_{\infty}^{2} \;,
\end{aligned}
\end{equation}
where $\cQ$ is defined in \eqref{eq:def Q modular} and $\tilde{q} = e^{-2\pi i / \ck\tau}$.

If one shifts the $\bdlt$'s by integers to the regime $-\im \frac{1}{\ck\tau} > \im\frac{m\bdlt_a}{\ck\tau} > 0$ before performing the modular transformations then the special functions on the right hand side above are exponentially suppressed in $n$. We denote 
\begin{equation}
[\bdlt]_{\tau} = \bdlt \mod 1 \;, \qquad -\im\frac{1}{\tau} > \im \frac{[\bdlt]_{\tau}}{\tau} > 0 \;.
\end{equation}
The special functions are now of order $O(\tilde{q}^n, \tilde{y}_a^n, (\tilde{q}^2/\tilde{y}_a)^n)$ where\footnote{We used $\dlt$ and not $\bdlt$ to define $y$, but it doesn't matter, as $\tilde y_a = \tilde q / e^{2\pi i (\Delta_1 + \Delta_2 - 1)/\ck\tau}$, so a definition using $\bdlt$ would amount to exchanging $\tilde y_3$ and $\tilde q / \tilde y_3$, both of which have magnitude smaller than 1 in our limit.} $\tilde{y}_a = e^{2\pi i [m\dlt_a]_{\ck\tau}/\ck\tau}$.

We now have to split into two cases, as in \cite{Aharony:2021zkr,Benini:2018ywd}:
\begin{equation}
\label{eq:def 1st 2nd case}
    \left[m\dlt_1 + m\dlt_2\right]_{\ck\tau} = \begin{cases} \left[m\dlt_1\right]_{\ck\tau} + \left[m\dlt_2\right]_{\ck\tau} & \text{\first\  case} \\
    \left[m\dlt_1\right]_{\ck\tau} + \left[m\dlt_2\right]_{\ck\tau} + 1 & \text{\second\  case}
    \end{cases}
\end{equation}
In order to compute the $b_N$ term in \eqref{eq:cI Hong Liu Contribution} up to exponential accuracy one needs to shift its second argument using \eqref{eq:shifting cG}, which changes $\dlt_a \to [\dlt_a]_{\ck\tau}$ in $b_N$. Using $[m\dlt_3] = 2\ck\tau - [m\dlt_1 + m\dlt_2] - 1$ we find that
\begin{equation}
b_N = \pm\frac{1}{\ck\tau} + \frac{1}{\ck\tau}\sum_{a=1}^3\cG\left(0;\frac{n[m\dlt_a]_{\ck\tau}}{\ck\tau},-\frac{n}{\ck\tau}\right)  \qquad \text{\first/\second \ case} \;,
\end{equation}
where the upper sign is for the first case and the lower for the second. For the first case
\begin{equation}
\begin{aligned}
    \log\mathcal{I}_{\left\{m,n,r\right\} } \xrightarrow{n \sim N \to \infty}& -\frac{\pi iN^{2}}{m}\frac{\left[m\dlt_{1}\right]_{\ck\tau} \left[m\dlt_{2}\right]_{\ck\tau} \left[m\dlt_3\right]_{\ck\tau}}{\check{\tau}^{2}} + \log N - \log \cI_{U(1)}\\
    &\qquad  + \frac{\pi i m}{12}  - \log\check{\tau} - \log\det\left(1+\frac{b_{N}^{-1}}{N}\mathcal{B}^{-1}\mathcal{C}\right) + O\left(\tilde{q}^n, \tilde{y}_a^n\right)\;,
\end{aligned}
\end{equation}
where $\left[m\dlt_3\right]_{\ck\tau} = 2\check{\tau} - \left[m\dlt_{1}+m\dlt_{2}\right]_{\ck\tau} - 1$. The determinant was evaluated in \eqref{eq:expansion determinant at large N m=1}, and using $\tilde{q} = \tilde{y}_1 \tilde{y}_2 \tilde{y}_3$
\begin{equation}
\begin{aligned}
    \det \left[\mathbf{1} + \frac{b_N^{-1}}{N}\cB^{-1}\cC \right]\Bigg|_{\first,m=1} &= \prod_{k=1}^{\infty}\left[1 - \sum_{a=1}^{3}\frac{\tilde{y}_{a}^{k} - \left(\tilde{q}/\tilde{y}_{a}\right)^{k}}{1-\tilde{q}^{k}}\right]^2 + O\left(e^{-n}\right) \\
    &= \prod_{k=1}^{\infty} \left[\frac{(1-\tilde y_1^k)(1-\tilde y_2^k)(1-\tilde y_3^k)}{\left(1-\tilde{q}^k\right)}\right]^2 + O\left(e^{-n}\right) \;,
\end{aligned}
\end{equation}
where the corrections are of order $O(\tilde y_a^n,(\tilde q / \tilde y_a)^n)$. Overall, the large $N$ limit of the $\{1,N,r\}$ Hong-Liu solution is
\begin{multline}
\label{eq:HL_1st_Large_N_m_1}
    \cI_{\{1,N,r\}} = \frac{e^{\frac{\pi i}{12}}}{\ck\tau \cI_{U(1)}} N e^{-\pi i N^2 \frac{\left[\dlt_1\right]_{\ck\tau}\left[\dlt_2\right]_{\ck\tau}\left[\dlt_3\right]_{\ck\tau}}{\ck\tau^2}} \prod_{k=1}^\infty \left[\frac{1 - \tilde q^k}{(1-\tilde y_1^k)(1-\tilde y_2^k)(1-\tilde y_3^k)}\right]^2 \\ \times \left[1 + O\left(\tilde q^N, \tilde y_a^N\right)\right] \,.
\end{multline}

For the second case, it is convenient to define $[m\dlt_a]^\prime_{\ck\tau} = [m\dlt_a]_{\ck\tau} + 1$. Then\footnote{In the case of even $N$ and odd $n$, the expression differs from that in \cite{Aharony:2021zkr} by $\pi i$.}
\begin{equation}
\begin{aligned}
    \log\mathcal{I}_{\left\{m,n,r\right\} } \xrightarrow{n \sim N \to \infty}& -\frac{\pi iN^{2}}{m}\frac{\left[m\dlt_{1}\right]^\prime_{\ck\tau} \left[m\dlt_{2}\right]^\prime_{\ck\tau} \left[m\dlt_3\right]^\prime_{\ck\tau}}{\check{\tau}^{2}} + \log N \\
    &\qquad - \log \cI_{U(1)} - \frac{\pi i m}{12} + \pi i + \pi i (n + N) \\
    &\qquad - \log\check{\tau} - \log\det\left(1+\frac{b_{N}^{-1}}{N}\mathcal{B}^{-1}\mathcal{C}\right) + O\left(\tilde{q}^n, \tilde{y}_a^n\right)\;,
\end{aligned}
\end{equation}
where $[m\dlt_3]_{\ck\tau}^\prime = 2\ck\tau - [m\dlt_1]_{\ck\tau}^\prime - [m\dlt_2]_{\ck\tau}^\prime + 1$. We find it convenient to define the fugacities $\tilde x_a = e^{-2\pi i [\dlt_a]^\prime_\tau} =  \tilde q / \tilde y_a$ which satisfy $\tilde x_1 \tilde x_2 \tilde x_3 = \tilde q$ and $|\tilde q| < |\tilde x_a| < 1$. Picking the bottom sign in \eqref{eq:expansion determinant at large N m=1} leaves us with the final expression for the deteminant in the second case
\begin{equation}
\begin{aligned}
    \det \left[\mathbf{1} + \frac{b_N^{-1}}{N}\cB^{-1}\cC \right]\Bigg|_{\second,m=1} &= \prod_{k=1}^{\infty}\left[1 + \sum_{a=1}^{3}\frac{\tilde{y}_{a}^{k} - \left(\tilde{q}/\tilde{y}_{a}\right)^{k}}{1-\tilde{q}^{k}}\right]^2 + O\left(e^{-n}\right) \\
    &= \prod_{n=1}^{\infty} \left[\frac{(1-\tilde x_1^k)(1-\tilde x_2^k)(1-\tilde x_3^k)}{1-\tilde{q}^k}\right]^2 + O\left(e^{-n}\right) \;,
\end{aligned}
\end{equation}
where the corrections are of order $O(\tilde x_a^n,(\tilde q / \tilde x_a)^n)$. Finally, for the second case the contribution of the $\{1,N,r\}$ Hong-Liu solution to the index is
\begin{multline}
\label{eq:HL_2nd_Large_N_m_1}
    \cI_{\{1,N,r\}} = -\frac{e^{-\frac{\pi i}{12}}}{\ck\tau \cI_{U(1)}} N e^{-\pi i N^2 \frac{\left[\dlt_1\right]_{\ck\tau}^\prime \left[\dlt_2\right]_{\ck\tau}^\prime \left[\dlt_3\right]_{\ck\tau}^\prime}{\ck\tau^2}} \prod_{k=1}^\infty \left[\frac{1 - \tilde q^k}{(1-\tilde x_1^k)(1-\tilde x_2^k)(1-\tilde x_3^k)}\right]^2 \\ \times \left[1 + O\left(\tilde q^N, \tilde x_a^N\right)\right] \,.
\end{multline}

\paragraph{The Cardy limit}
In the Cardy limit $\tau \to 0$ the dominant contribution to the index comes from the $\{m,n,r\} = \{1,N,0\}$ solution \cite{Benini:2018ywd,GonzalezLezcano:2020yeb,Cabo-Bizet:2019eaf}, where the holonomies approach $u_i = 0$. In order to get a good expansion\footnote{Note that we don't need to expand in large $N$ before taking the Cardy limit - terms that are suppressed in $N$ are also suppressed at high temperatures.} in $1/\tau$ one should use the modular transformations \eqref{eq:modular transforms special functions} on $\cI_{U(1)}$ as well as on the rest of \eqref{eq:cI Hong Liu Contribution},
\begin{equation}
\begin{aligned}
\log\cI_{U\left(1\right)} &= \frac{\pi i}{2} - \log\tau - \frac{\pi i}{6}\left(\tau+\frac{1}{\tau}\right)+\log\left(\tilde{q};\tilde{q}\right)_{\infty}^{2} \\
&\quad - \sum_{a=1}^{3}\left[\pi i\cQ\left([\dlt_{a}]_\tau;\tau;\tau\right) - \log\theta_{0}\left(\frac{[\dlt_{a}]_\tau}{\tau};-\frac{1}{\tau}\right) + \sum_{k=0}^{\infty}\log\frac{\psi\left(\frac{k+1+[\dlt_{a}]_\tau}{\tau}\right)}{\psi\left(\frac{k-[\dlt_{a}]_\tau}{\tau}\right)}\right] \;,
\end{aligned}
\end{equation}
so the index in the Cardy limit is
\begin{equation}
\log \cI \xrightarrow{\tau \to 0} \begin{cases} - \pi i(N^{2}-1)\frac{[\dlt_{1}]_\tau[\dlt_{2}]_\tau[\dlt]_\tau}{\tau^{2}} + \log N  + O\left(\tilde q, \tilde y_a\right)  & \text{\first\ case} \\
- \pi i(N^{2}-1)\frac{[\dlt_{1}]_\tau^\prime [\dlt_{2}]_\tau^\prime [\dlt]_\tau^\prime}{\tau^{2}} + \log N  + O\left(\tilde q, \tilde x_a\right)  & \text{\second\ case}
\end{cases}\;.
\end{equation}
When $\dlt_1 = \dlt_2 = \dlt_3 = \frac{2\tau - 1}{3}$ the expression agrees with \cite{ArabiArdehali:2021nsx,Cassani:2021fyv,Cabo-Bizet:2019osg,Kim:2019yrz,GonzalezLezcano:2020yeb}. Specifically, note that the prefactor $N^2$ changes in this limit to $N^2-1$ and that the $\log \tau$ term cancels out, both are necessary implications of the EFT argument of \cite{Cassani:2021fyv}. But we can do better. Given the exact expression for the determinant term \eqref{eq:expansion determinant at large N m=1} and for the special functions involved, we can systematically write down the leading corrections to $\cI_{\{1,N,0\}}$. The rather simple expression for the first case is
\begin{equation}
\boxed{\begin{aligned}
\cI_{\{1,N,0\}}\big|_{\first} \xrightarrow{\tau \to 0}
\frac{e^{\frac{\pi i}{12}}}{\tau \cI_{U(1)}} N e^{-\pi i N^2\frac{ [\dlt_1]_\tau [\dlt_2]_\tau [\dlt_3]_\tau}{\tau^2}}
\prod_{k=1}^\infty &\left[\frac{1-\tilde{q}^k}{\left(1 - \tilde{y}_1^k\right)\left(1 - \tilde{y}_2^k\right)\left(1 - \tilde{y}_3^k\right)}\right]^2 \\
&\qquad \times \left(1 + O\left(\tilde y_a^N, \tilde q^N, (\tilde q / \tilde y_a)^N\right)\right)
\end{aligned}}
\end{equation}
while for the second case it is
\begin{equation}
\boxed{\begin{aligned}
\cI_{\{1,N,0\}}\big|_{\second} \xrightarrow{\tau \to 0} -\frac{e^{-\frac{\pi i}{12}}}{\tau \cI_{U(1)}} N e^{-\pi i N^2 \frac{[\dlt_1]_\tau^\prime [\dlt_2]_\tau^\prime [\dlt_3]_\tau^\prime}{\tau^2}} \prod_{k=1}^\infty &\left[\frac{1-\tilde{q}^k}{\left(1 - \tilde{x}_1^k\right)\left(1 - \tilde{x}_2^k\right)\left(1 - \tilde{x}_3^k\right)}\right]^2 \\
&\qquad  \times \left(1 + O\left(\tilde x_a^N, \tilde q^N, (\tilde q / \tilde x_a)^N\right)\right)
\end{aligned}}
\end{equation}
where $\tilde y_a = e^{2\pi i [\dlt]_a/\tau}$, $\tilde q = e^{-2\pi i/\tau}$, $x_a = \tilde q / \tilde y_a$. These formulas receive corrections at order $O(\tilde x^N, \tilde y_a^N, \tilde q^N)$. Note that $\frac{e^{\frac{\pi i}{12}}}{\tau \cI_{U(1)}}$ changes the exponent to $N^2 - 1$ and adds $O(\tilde q, \tilde y_a)$ terms to the action, so it should also be considered if one wants to expand the order $O(e^{-1/\tau})$ terms in the Cardy limit order by order. $\cI_{U(1)}$ is defined in \eqref{eq:def integrand}.

Assuming no other Bethe Ansatz configuration contributes in this limit, these are also the leading asymptotics of the index itself. This assumption applies for all other HL configurations, as they are smaller by $O(e^{-N^2/\tau^2})$. We assume that in the Cardy limit other contributions are also suppressed by at least $O(e^{-N/\tau})$.

\paragraph{Acknowledgment}
The author thanks Erez Urbach and Tal Sheaffer for useful discussions, and is grateful to Francesco Benini and Ofer Aharony for many useful conversations and for comments on the manuscript.
The work of the author was supported in part by an Israel Science Foundation (ISF) center for excellence grant (grant number 2289/18), by ISF grant no. 2159/22, by Simons Foundation grant 994296 (Simons Collaboration on Confinement and QCD Strings), by grant no. 2018068 from the United States-Israel Bi- national Science Foundation (BSF), by the Minerva foundation with funding from the Federal German Ministry for Education and Research, by the German Research Foundation through a German-Israeli Project Cooperation (DIP) grant “Holography and the Swampland”, and by a research grant from Martin Eisenstein. The author would like to thank SISSA for its hospitality, funded by ERC-COG grant NP-QFT No. 864583.
The author also thanks the Simons Center for Geometry and Physics, Stony Brook University for organizing the ``Supersymmetric Black Holes, Holography and Microstate Counting'' workshop which helped facilitate this paper.

\appendix

\section{The superconformal index of $\mathcal{N} = 4$ SYM}
\label{app:Index def}
Using $\cN=1$ language, $\cN=4$ SYM theory consists of a vector multiplet and three chiral multiplets in the adjoint representation. Its R-symmetry is $SU(4)_R$, whose Cartan is $U(1)^3$. We pick generators $R_{1,2,3}$, each giving R-charge 2 to a single chiral multiplet and zero to the other two, in a symmetric way. Local operators in the theory (and likewise states on $S^3$) are labeled by two half-integer angular momenta $J_{1,2}$, each rotating an $\bR^2 \subset \bR^4$ in which $S^3$ is embedded. The fermion number is defined as $F = 2J_1$. Note that all fields in the theory (and thus all states) have integer charges under $R_{1,2,3}$ and obey
\begin{equation} \label{eq:modone}
F = 2J_{1,2} = R_{1,2,3} \pmod{2} \;.
\end{equation}

The superconformal index of ${\cal N}=4$ SYM theory counts, with sign, 1/16-BPS states on $S^3$ preserving one complex supercharge $\cQ$, which we choose to be associated with a specific $U(1)_R$ symmetry the generator $r=\frac{1}{3}(R_1+R_2+R_3)$. The index can also keep track of some combinations of the R-charges and the two angular momenta $J_{1,2}$. It is useful to introduce two flavor generators $\fq_{1,2} = \frac12 (R_{1,2} - R_3)$ that commute with the supercharge and with $r$. 
The superconformal index \cite{Romelsberger:2005eg, Kinney:2005ej} is then defined by the trace%
\footnote{Often the powers in the index are written as $p^{J_1+\frac12 r} \, q^{J_2+\frac12 r}$. Compared to this convention, we have swallowed a power of $(pq)^{1/3}$ into $y_1$ and $y_2$ in order to obtain a single-valued function. The relation of our variables to those of \cite{Kinney:2005ej} is $p = t^3y \big|_\text{there}$, $q = t^3/y \big|_\text{there}$, $y_1 = t^2v \big|_\text{there}$, $y_2 = t^2 w/v \big|_\text{there}$.}
\begin{equation}
\label{eq:index Hamiltonian definition}
\cI(p,q,y_1, y_2) = \Tr\left[ (-1)^F \, e^{-\beta\{\cQ, \bar\cQ \}} \, p^{J_1 + \frac12 R_3} \, q^{J_2 + \frac12 R_3} \, y_1^{\fq_1} \, y_2^{\fq_2} \right]
\end{equation}
over the Hilbert space on $S^3$, where we normalized the vacuum contribution to 1. Here $p$, $q$, $y_{1,2}$ are fugacities, and it is convenient to introduce chemical potentials $\sigma$, $\tau$, $\Delta_{1,2}$ such that
\begin{equation} \label{eq:fugacities}
p = e^{2\pi i \sigma} \;,\qquad\qquad q = e^{2\pi i \tau} \;,\qquad\qquad y_{1,2} = e^{2\pi i \Delta_{1,2}} \;.
\end{equation}
The index is well-defined for $|p|, |q|<1$, namely for $\im(\tau), \im(\sigma) > 0$. Given \eqref{eq:modone},
the index is a single-valued function of the fugacities \eqref{eq:fugacities} - it is periodic under integer shifts of the chemical potentials $\sigma$, $\tau$, $\Delta_1$ and $\Delta_2$. By standard arguments 
\cite{Witten:1982df}, the index only counts states annihilated by $\cQ$ and $\bar{\cQ}$ and is thus independent of $\beta$.

In addition to the index, one could also compute the thermal partition function of the theory on $S^1 \times S^3$ in the presence of general chemical potentials
\begin{equation}
Z_{S^1\times S^3} = \Tr\left[ e^{-\beta H + \sum_{i=1}^2\beta \Omega_i J_i + \frac{1}{2}\sum_{a=1}^3\beta \Phi_a R_a} \right] \;.
\end{equation}
When the chemical potentials satisfy the constraint \cite{Cabo-Bizet:2018ehj}
\begin{equation}
\label{eq:chem pot const}
\beta\left(1 + \Omega_1 + \Omega_2 - \Phi_1 - \Phi_2 - \Phi_3\right) = 2\pi i n \;, \qquad n \in \bZ \;,
\end{equation}
there is a well defined complex Killing spinor on this space. The Killing spinor is periodic around the thermal cycle when $n$ is even, and it is anti-periodic when $n$ is odd, and so are the other fermions in the theory. The existence of the Killing spinor implies the existence of a complex supercharge $\cQ$ on the manifold, whose algebra is \cite{Closset:2013vra}
\begin{equation}
    \{\cQ,\bar\cQ\} = H - J_1 - J_2 - \frac{3}{2} r\;.
\end{equation}
Combining this relation with the constraint \eqref{eq:chem pot const}, the thermal partition function becomes
\begin{multline}
Z_{S^1 \times S^3} = \Tr\Big[(-1)^{(n+1)F} e^{\pi i n R_3} \exp\Big\{-\beta\{\cQ,\bar \cQ\} + 2\pi i \tau \left(J_1 + \frac{1}{2}R_3 \right) \\ + 2\pi i \sigma \left(J_2 + \frac{1}{2}R_3\right) + \pi i \Delta_1(R_1 - R_3) + \pi i \Delta_2(R_2 - R_3)\Big\} \Big] \;,
\end{multline}
the additional $(-1)^{(n+1)F}$ coming from the periodicity of the fermions, and we defined 
\begin{equation}
\tau = \frac{\beta(\Omega_1 - 1)}{2\pi i } \;,\qquad \sigma = \frac{\beta(\Omega_2 - 1)}{2\pi i} \;, \Delta_a = \frac{\beta(\Phi_a - 1)}{2\pi i} \;,
\end{equation}
for convenience. The constraint \eqref{eq:chem pot const} is now
\begin{equation}
\tau + \sigma - \Delta_1 - \Delta_2 - \Delta_3 = n \;, \qquad n\in\bZ \;.
\end{equation}
Further simplification arises after applying \eqref{eq:modone}, and the partition function turns out to be 
\begin{equation}
Z_{S^1 \times S^3} = \Tr\left[(-1)^F e^{-\beta\{\cQ,\bar \cQ\} + 2\pi i \tau \left(J_1 + \frac{1}{2}R_3 \right)  + 2\pi i \sigma \left(J_2 + \frac{1}{2}R_3\right) + 2\pi i \Delta_1 \fq_1 + 2\pi i \Delta_2\fq_2} \right] \;,
\end{equation}
which is precisely the index \eqref{eq:index Hamiltonian definition} up to the normalization of the vacuum. 

The index is often written when $\tau, \sigma$ couple to $J_{1,2} + \frac{1}{2}r$, instead as in \eqref{eq:index Hamiltonian definition}. A redefinition of the flavor chemical potentials, $\xi_{1,2} = \Delta_{1,2} + \frac{\tau + \sigma}{3}$, transforms the index to that form.

We note that for the particular choice $\Delta_{1,2} = \frac{2\tau - 1}{3}$, which fits $n=1$, the partition function equals (up to the vacuum contribution) to the index used in \cite{Cassani:2021fyv}, also termed the R-charge index or the superconformal index on the second sheet,
\begin{equation}
    Z_{S^1\times S^3} \sim \cI_R = \Tr\left[(-1)^R p^{J_1 + \frac{1}{2}r} q^{J_2 + \frac{1}{2}r}\right] \;.
\end{equation}

\section{Simplifying the Hong-Liu contribution}
\label{app:simplifying HL contribution}
Let us compute the contribution of the $\{m,n,r\}$ Hong-Liu solution \eqref{eq:Hong Liu solutions} to the index. Each such solution contributes as 
\begin{equation}
    \cI_{\{m,n,r\}} = N \cdot N! \cdot \frac{\cI_{U(1)}^{N-1}}{N!} \cdot \cZ(u_{\{m,n,r\}};\bdlt,\tau) \cdot H^{-1}\left(u_{\{m,n,r\}};\bdlt,\tau\right) \;,
\end{equation}
Here $\cZ$, $H$, and $\cI_{U(1)}$ are defined in \eqref{eq:Index integral formula}, \eqref{eq:Jacobian def}. The factor $N \cdot N!$ takes into account the multiplicity of each solution - a factor of $N!$ comes from the Weyl group action that permutes the eigenvalues $\{u_i\}$, and the factor of $N$ comes from the inequivalent choice of points on the torus that give rise to the same $u_{ij}$'s, and correspond to the action of the center symmetry as it shifts the center of mass of the first $N-1$ holonomies by $\frac{\ell_1 \tau + \ell_2}{N}$, where $\ell_1 = 0,\dots,m-1$ and $\ell_2 = 0,\dots,n-1$. 

We also remind that in our notation $\ck\tau = m\tau + r$ for the $\{m,n,r\}$ solution, and we will use two different conventions for the chemical potentials, $\bdlt_{1,2} = \dlt_{1,2}$ and $\bdlt_3 = \bdlt_1 + \bdlt_2$ or $\dlt_3 = 2\tau - \dlt_1 - \dlt_2 - 1$, such that $\dlt_3 = 2\tau - \bdlt_3 - 1$. Some computations are easier using the barred notation and some using the unbarred one. The paper \cite{Aharony:2021zkr} uses exclusively unbarred notation.

\subsection{Computing $\cZ$}
We need to compute
\begin{equation}
    \log \mathcal{Z}(u;\bdlt,\tau) = \gamma_{\bdlt_1} + \gamma_{\bdlt_2} - \gamma_{\bdlt_1+\bdlt_2}  - \gamma_0 \;, \qquad \gamma_\bdlt \equiv \sum_{i\neq j}^{N-1}\log \wt\Gamma(u_{ij} + \bdlt;\tau,\tau)\;.
\end{equation}
For any function $f$, $\sum_{i\neq j=0}^{M-1} f(u_i - u_j) = \sum_{i,j=0}^{M-1} f(u_i - u_j) - M f(0)$, and thus summation over regular shifts along the two cycles of the $(1,\tau)$ torus is given by\footnote{One can check that the series expansions are valid for some domain for the arguments, and then analytically continued to generic argument. For the first identity the domain is $0<\im(\bdlt)<2\im(\tau)$, while for the second the domain is $0<\im\left(\frac{n-1}{n}\tau\right)<\im(\bdlt)<\im\left(\frac{n+1}{n}\tau\right)$.}
\begin{multline}
    \sum_{\hat\imath\neq \hat\jmath=0}^{m-1} \log\wt\Gamma\left(u + \frac{\hat\imath-\hat\jmath}{m};\tau,\tau\right) + m \log\wt\Gamma(u;\tau,\tau)
    = \sum_{\hat\imath,\hat\jmath=0}^{m-1} \log\wt\Gamma\left(u + \frac{\hat\imath-\hat\jmath}{m};\tau,\tau\right) 
    \\ = \sum_{\hat\imath,\hat\jmath=0}^{m-1}\sum_{\ell=1}^\infty \frac{1}{\ell}\frac{z^\ell e^{2\pi i\ell(\hat\imath - \hat\jmath)/m} - q^{2\ell} z^{-\ell} e^{-2\pi i \ell(\hat\imath - \hat\jmath)/m}}{(1-q^\ell)^2}
    = m^2\sum_{\ell=1}^\infty \frac{1}{m\ell}\frac{z^{m\ell} - q^{2m\ell} z^{-m\ell} }{(1-q^{m\ell})^2} 
    \\ = m\log\wt\Gamma(mu;m\tau,m\tau) \;,
\end{multline}
and
\begin{multline}
    \sum_{\hat\imath\neq \hat\jmath}^{n-1} \log\wt\Gamma\left(u + \frac{\hat\imath-\hat\jmath}{n}\tau;\tau,\tau\right) + n \log\wt\Gamma(u;\tau,\tau)
    = \sum_{\hat\imath,\hat\jmath=0}^{n-1} \log\wt\Gamma\left(u + \frac{\hat\imath-\hat\jmath}{n}\tau;\tau,\tau\right) \\
    = \sum_{\hat\imath,\hat\jmath=0}^{n-1}\sum_{\ell=1}^\infty \frac{1}{\ell}\frac{z^\ell q^{(\hat\imath - \hat\jmath)\ell/n} - q^{2\ell} z^{-\ell} q^{-(\hat\imath - \hat\jmath)\ell/n}}{(1-q^\ell)^2} 
    = \sum_{\ell=1}^\infty \frac{1}{\ell}\frac{z^\ell q^{-\ell+\frac{\ell}{n}}\frac{(1-q^\ell)^2}{(1-q^{\frac{\ell}{n}})^2} - q^{2\ell} z^{-\ell} q^{-\ell+\frac{\ell}{n}}\frac{(1-q^\ell)^2}{(1-q^{\frac{\ell}{n}})^2}}{(1-q^\ell)^2} \\
    = \sum_{\ell=1}^\infty \frac{1}{\ell}\frac{z^\ell q^{-\ell+\frac{\ell}{n}} - q^{\ell+\frac{\ell}{n}}z^{-\ell}}{(1-q^{\frac{\ell}{n}})^2}
    = \log\wt\Gamma\left(u - (n-1)\frac{\tau}{n};\frac{\tau}{n},\frac{\tau}{n}\right) \;,
\end{multline}
The argument of the resulting elliptic Gamma function can be shifted,
\begin{equation}
\begin{aligned}
    \log\wt\Gamma&\left(u - (n-1)\frac{\tau}{n};\frac{\tau}{n},\frac{\tau}{n}\right) - \log\wt\Gamma\left(u;\frac{\tau}{n},\frac{\tau}{n}\right)\\
    &= -\sum_{k=1}^{n-1}\log\theta_0\left(u - k\frac{\tau}{n};\frac{\tau}{n}\right) \\
    &= - (n-1)\log\theta_0\left(u;\frac{\tau}{n}\right) - \sum_{k=1}^{n-1}\left[\pi i k + \pi i k \left(2u - (k+1)\frac{\tau}{n}\right)\right] \\
    &= - (n-1)\log\theta_0\left(u;\frac{\tau}{n}\right) - \frac{\pi i (n-1)}{3}(3 n u - (n+1)\tau) - \pi i (n-1)\;.
\end{aligned}
\end{equation}

Overall, for the $\{m,n,r\}$ HL solution and when $\bdlt\neq 0$ we find
\begin{equation}
\begin{aligned}
    \gamma_\bdlt &= \sum_{\hat k\neq\hat\ell}^{n-1} \sum_{\hat\imath,\hat\jmath=0}^{m-1}\log \wt\Gamma\left(\frac{\hat\imath - \hat\jmath}{m} + \frac{\hat k - \hat\ell}{n}\left(\tau + \frac{r}{m}\right) + \bdlt;\tau,\tau\right) \\
    &\quad + n\sum_{\hat\imath\neq\hat\jmath}^{m-1}\log\wt\Gamma\left(\frac{\hat\imath - \hat\jmath}{m}+\bdlt;\tau,\tau\right) \\
    &= m\sum_{\hat k\neq\hat\ell}^{n-1} \log \wt\Gamma\left(\frac{\hat k - \hat\ell}{n}\ck\tau + m\bdlt;\ck\tau,\ck\tau\right) + N\log\frac{\wt\Gamma\left(m\bdlt;\ck\tau,\ck\tau\right)}{\wt\Gamma\left(\bdlt;\tau,\tau\right)} \\
    &= m\log\wt\Gamma\left(m\bdlt - (n-1)\frac{\ck\tau}{n} ; \frac{\ck\tau}{n},\frac{\ck\tau}{n}\right) - N\log\wt\Gamma\left(\bdlt;\tau,\tau\right) \\
    &= m\log\wt\Gamma\left(m\bdlt; \frac{\ck\tau}{n},\frac{\ck\tau}{n}\right) - (N-m)\log\theta_0\left(m\bdlt;\frac{\ck\tau}{n}\right) - N\log\wt\Gamma\left(\bdlt;\tau,\tau\right) \\
    & \qquad - \frac{\pi i (N-m)}{3}(3 N\bdlt - (n+1)\frac{\ck\tau}{n}) - \pi i (n-1)\;,
\end{aligned}
\end{equation}
where $\ck\tau = m\tau + r$. The limit $\bdlt \to 0$ can now be taken as well. This is most conveniently done by expressing the elliptic Gamma and theta functions as an infinite product
\begin{equation}
\begin{aligned}
    \lim_{u\to 0} \log\Gamma\left(u;\tau,\tau\right) &= \lim_{z\to 1} \log\left[\frac{\prod_{k,\ell=0}^\infty\left(1-q^{k+1}q^{\ell+1}z^{-1}\right)}{\prod_{k,\ell=0}^\infty\left(1-q^kq^\ell z\right)}\right] \\
    &= \lim_{z\to 1}\log \left( \frac{1}{1-z} \right) + 2\log\left[\prod_{\ell=1}^\infty\frac{1}{1-q^\ell}\right] \\
    &= \lim_{z\to 1}\log\left( \frac{1}{1-z} \right) - 2\log(q;q)_\infty \;,
\end{aligned}
\end{equation}
and
\begin{equation}
\begin{aligned}
    \lim_{u\to 0} \log\theta_0\left(u;\tau\right) &= \lim_{z\to 1} \log\left[\prod_{k=0} (1 - z q^k)(1-  z^{-1}q^{k+1})\right] \\
    &= \lim_{z\to 1} \log(1-z) + \log\left[\prod_{k=1} (1 - q^k)(1 - q^k)\right] \\
    &= \lim_{z\to 1} \log(1-z) + 2\log(q;q)_\infty
\end{aligned}
\end{equation}
so
\begin{equation}
    \gamma_0 = \lim_{\bdlt\to 0} \gamma_{\bdlt} = \frac{\pi i (N-m)(n+1)}{3}\frac{\ck\tau}{n} - \pi i (n-1) - N\log m + 2N\log\frac{\left(q;q\right)_\infty}{\left(\ck q^{\frac{1}{n}};\ck q^{\frac{1}{n}}\right)_\infty}  \;.
\end{equation}
Finally, we can put it all together to compute the total contribution $\cZ$ with $\ck q = e^{2\pi i \ck\tau}$
\begin{equation}
\boxed{
\label{eq:cZ HL solutions}
\begin{aligned}
    \cZ\big(u_{\{m,n,r\}},& \bdlt_1, \bdlt_2, \tau\big) \\
    &= \sum_{a=1}^3 \eta_a \Big[m\log\wt\Gamma\left(m\bdlt_a;\frac{\ck\tau}{n},\frac{\ck\tau}{n}\right) + (m-N)\log\theta_0\left(m\bdlt_a;\frac{\ck\tau}{n}\right) \\ 
    &\qquad \qquad - N\log\wt\Gamma(\bdlt_a;\tau,\tau) \Big] + N\log m - 2N\log\frac{\left(q,q\right)_\infty}{\left(\ck q^{\frac{1}{n}},\ck q^{\frac{1}{n}}\right)_\infty}
\end{aligned}}
\end{equation}
where we have denoted $\bdlt_3 = \bdlt_1 + \bdlt_2$ and $\eta_a = (1, 1, -1)$. 

\subsection{The Jacobian $H$}
The Jacobian can be written as (see appendix B.3 in \cite{Aharony:2021zkr}) 
\begin{equation}
    \log H = N\log N+\left(N-1\right)\log b_{N} + \log\det\left(1+\frac{b_{N}^{-1}}{N}\cB^{-1}\cC\right) \;,
\end{equation}
where
\begin{equation}
\label{eq:matrices defs}
\begin{aligned}
    b_N &= -4 + \frac{1}{\ck\tau}\sum_{a=1}^3\left[2m\dlt_a + 1 + \cG\left(0,\frac{N\dlt_a}{\ck\tau};-\frac{n}{\ck\tau}\right)\right] \;,
    \\
    \cB_{ij} &= \bdlt_{ij}+1 \;, \qquad \cB^{-1}_{ij} = \bdlt_{ij} - \frac{1}{N} \;, \\
    \cC_{ij} &= \sum_{a=1}^{3}\left[\cG\left(u_{iN},\dlt_{a};\tau\right) + \cG\left(u_{jN}, \dlt_{a};\tau\right) - \cG\left(u_{ij},\dlt_{a};\tau\right) - \cG\left(0,\dlt_{a};\tau\right)\right] \;.
\end{aligned}
\end{equation}
We will now compute $\log\det\left(1+\frac{b_{N}^{-1}}{N}\cB^{-1}\cC\right)$ by diagonalizing the $(N-1) \times (N-1)$ matrix $\cX = \cB^{-1}\cC$. Only the expression for $m = 1$ will be reproduced here\footnote{The more general case $n \gg m > 1$ is interesting and relevant for the orbifold saddles. While the eigenvalues for that case can be found as a formal power series, it is not one with a good large $N$ limit, in the sense that it contains powers of $q^{1/n}$. Perhaps some modular properties could be used to get a good large $N$ limit, but we were not able to do that.}. However, we will restrict to $m=1$ only when needed, at \eqref{eq:eigenvalue equation m=1}. Let's begin. Using (B.26) and (B.29) in \cite{Aharony:2021zkr}
\begin{multline}
    \Upsilon \equiv \sum_{j=1}^N \cG\left(u_{j_1j},\bdlt;\tau\right) = \sum_{\hat\jmath=0}^{m-1}\sum_{k=0}^{n-1}\cG\left(\frac{\hat\jmath_1-\hat\jmath}{m}+\frac{k_1-k}{n}\frac{\ck\tau}{m},\bdlt;\tau\right) \\
    = \frac{N}{\ck\tau}\left[2m\bdlt_a + 1 - \ck\tau + \cG\left(0,\frac{N\bdlt}{\ck\tau};-\frac{n}{\ck\tau}\right)\right]
    = m-N + m\cG\left(0,m\bdlt;\frac{\ck\tau}{n}\right)
\end{multline}
is independent of $j_1$. Suppressing the $\tau$ argument of $\cG$ for brevity,
\begin{multline}
    \sum_{\ell=1}^{N-1}\cC_{\ell j} = \sum_{\ell=1}^{N-1}\sum_{a=1}^{3}\left[\cG\left(u_{\ell N},\dlt_{a}\right)+\cG\left(u_{jN},\dlt_{a}\right)-\cG\left(u_{\ell j},\dlt_{a}\right)-\cG\left(0,\dlt_{a}\right)\right] \\
    = \sum_{a=1}^{3}\Big[\Upsilon-\cG\left(u_{NN},\dlt_{a}\right)+\left(N-1\right)\cG\left(u_{jN},\dlt_{a}\right) - \left(\Upsilon-\cG\left(u_{Nj},\dlt_{a}\right)\right)-\left(N-1\right)\cG\left(0,\dlt_{a}\right)\Big] \\
    = N\sum_{a=1}^{3}\left[\cG\left(u_{jN},\dlt_{a}\right)-\cG\left(0,\dlt_{a}\right)\right] \;.
\end{multline}
where we used the fact that $\cG(u,\dlt)$ is an even function of $u$, and so
\begin{equation}
    \cX_{ij} = \cC_{ij} - \frac{1}{N} \sum_{\ell=1}^{N-1} \cC_{\ell j} = \sum_{a=1}^{3} \left[\cG\left(u_{iN},\dlt_{a}\right) - \cG\left(u_{ij},\dlt_{a}\right)\right]
\end{equation}
Now let's find the eigenvalues $\lambda$ and eigenvectors $v$ of $\cX$. Denote $\bar{v}=\sum_{j=1}^{N-1}v_{j}$. The eigenvalue equations are
\begin{equation*}
    \lambda v_{i} = \sum_{j=1}^{N-1} \cX_{ij}v_{j} = \sum_{a=1}^{3} \left[\cG\left(u_{iN},\dlt_{a}\right)\bar{v} - \sum_{j=1}^{N-1} \cG\left(u_{ij},\dlt_{a}\right)v_{j}\right] \;,\qquad i=1,\dots,N-1 \;,
\end{equation*}
Now define a vector of length $N$, $w = \left(v_{1},v_{2},\dots,v_{N-1},-\bar{v}\right)$, for which
\begin{equation}
    \lambda w_{i} = -\sum_{j=1}^{N}\sum_{a=1}^{3} \cG\left(u_{ij},\dlt_{a}\right)w_{j} \;, \qquad i=1,\dots,N-1 \;.
\end{equation}
By summing all the equations and multiplying by $-1$ we also find the equations for $w_N$
\begin{multline*}
\lambda w_{N} = -\sum_{a=1}^{3} \left[\Upsilon\bar{v}-\cG\left(0,\dlt_{a}\right)\bar{v} - \sum_{j=1}^{N-1} \left(\Upsilon-\cG\left(u_{jN},\dlt_{a}\right)\right)v_{j}\right] \\
= -\sum_{j=1}^{N-1}\sum_{a=1}^{3} \left[\cG\left(u_{jN},\dlt_{a}\right)v_{j}-\cG\left(0,\dlt_{a}\right)\bar{v}\right] 
= -\sum_{j=1}^{N}\sum_{a=1}^{3}\cG\left(u_{jN},\dlt_{a}\right)w_{j}
\end{multline*}
so the eigenvalue equations are (note the range of $i$)
\begin{equation}
    \lambda w_{i} = -\sum_{j=1}^{N}\sum_{a=1}^{3} \cG\left(u_{ij},\dlt_{a};\tau\right)w_{j} \;, \qquad i=1,\dots,N \;.
\end{equation}
Writing the indices using the $u_i \equiv u_{\hat\imath \hat\ell}$ notation reduces us to the form
\begin{equation}
\label{eq:eigenvalue equation w}
    \lambda w_{\hat\imath\hat\ell} = -\sum_{\hat\jmath=0}^{m-1}\sum_{\hat k=0}^{n-1}  f(\hat\imath-\hat\jmath, \hat\ell - \hat k) w_{\hat\jmath\hat\ell} \;, \qquad f(\hat\jmath,\hat k) = \sum_{a=1}^3 \cG\left(\frac{\hat\jmath}{m} + \frac{\hat k}{n}(\tau + \frac{r}{m}),\dlt_a;\tau\right) \;.
\end{equation}
These equations can be solved by a discrete Fourier transform, defined by
\begin{equation}
    \wt w_{b,c} = \sum_{\hat\jmath=0}^{m-1}\sum_{\hat k=0}^{n-1} w_{\hat\jmath\hat k}e^{-2\pi i \hat\jmath b / m} e^{-2\pi i \hat k c / n} \;,
\end{equation}
such that the eigenvalue equations become\footnote{The Fourier transform of the discrete convolution is \begin{equation*}
    \sum_{\hat k=0}^{N-1} \sum_{\hat\jmath=0}^{N-1} f(\hat k - \hat\jmath) g(\hat\jmath) e^{-2\pi i \hat k a / N} = \sum_{\hat k=0}^{N-1} \sum_{\hat\jmath=0}^{N-1} \sum_{b=0}^{N-1} \frac{\tilde f(b)}{N} e^{2\pi ib(\hat k - \hat\jmath) / N} g(\hat\jmath)e^{-2\pi i \hat k a / N}  \\ = \sum_{\hat\jmath=0}^{N-1} \tilde f(a) e^{-2\pi i a \hat\jmath / N} g(\hat\jmath) = \tilde f(a) \tilde g(-a) \;.
\end{equation*}}
\begin{equation}
\label{eq:eigenvalues as Fourier trans}
    \lambda\wt w_{b,c} = - \wt f_{b,c} \wt w_{b,c} \;.
\end{equation}
The eigenvalues are thus $\lambda_{b,c} = -\wt f_{b,c}$ for $b = 0,\dots,m-1$, $c = 0,\dots,n-1$, besides $\wt f_{0,0}$, as $\wt w_{0,0} = \sum_{j=1}^{N} w_j = \sum_{j=1}^{N-1} v_j - \bar v = 0$, the equation for $a=b=0$ is always satisfied, and there are indeed only $N-1$ eigenvalues as expected for our $(N-1) \times (N-1)$ matrix $\cX$.

% Since $\cG(u,\dlt;\tau)$ is an even function of $u$, $\wt f_{b,c} = \wt f_{m-b, n-c}$, and the eigenvalues come in pairs (besides maybe one eigenvalue if $N$ is even). 

\paragraph{The case $m = 1$}
Let us now restrict ourselves to the case $m=1, \ck\tau = \tau + r$. The eigenvalues are
\begin{equation}
\label{eq:eigenvalue equation m=1}
    \lambda_{n} = -\sum_{j=0}^{N-1}\sum_{a=1}^{3}\cG\left(\frac{j\ck\tau}{N},\dlt_{a};\ck\tau\right)e^{2\pi ijn/N}\;,\qquad n=1,\dots,N-1
\end{equation}
By using the modular properties of $\cG$, $ \cG\left(u,\dlt;\ck\tau\right) = \frac{1}{\ck\tau}\cG\left(\frac{u}{\ck\tau},\frac{\dlt}{\ck\tau};-\frac{1}{\ck\tau}\right)-1-\frac{2\dlt}{\ck\tau}-\frac{1}{\ck\tau}$ and the series expansion $\cG\left(u,\dlt;\tau\right) = \sum_{\ell=1}^{\infty}\frac{\left(z^{\ell}+z^{-\ell}\right)\left(y^{\ell}-\left(q/y\right)^{\ell}\right)}{1-q^{\ell}}$ we note that for $\tilde{q}=e^{-2\pi i/\ck\tau}$, $\tilde{y}_{a}=e^{-2\pi i\dlt_{a}/\ck\tau}$
\begin{equation}
\label{eq:explicit eigenvalues m=N}
\begin{aligned}
    -\lambda_{n}\ck\tau &= \sum_{a=1}^{3}\sum_{j=0}^{N-1}\sum_{\ell=1}^{\infty}\frac{\left(e^{2\pi ij\ell/N}+e^{-2\pi ij\ell/N}\right)\left(\tilde{y}_{a}^{\ell}-\left(\tilde{q}/\tilde{y}_{a}\right)^{\ell}\right)}{1-\tilde{q}^{\ell}}e^{2\pi ijn/N} \\ 
    &= \sum_{a=1}^{3}\sum_{\ell=1}^{\infty}\sum_{j=0}^{N-1}\frac{\left(e^{2\pi ij\left(\ell+n\right)/N}+e^{-2\pi ij\left(\ell-n\right)/N}\right)\left(\tilde{y}_{a}^{\ell}-\left(\tilde{q}/\tilde{y}_{a}\right)^{\ell}\right)}{1-\tilde{q}^{\ell}} \\ 
    &= N\sum_{a=1}^{3}\left[\sum_{k=1}^{\infty}\frac{\tilde{y}_{a}^{kN-n}-\left(\tilde{q}/\tilde{y}_{a}\right)^{kN-n}}{1-\tilde{q}^{kN-n}}+\sum_{k=0}^{\infty}\frac{\tilde{y}_{a}^{kN+n}-\left(\tilde{q}/\tilde{y}_{a}\right)^{kN+n}}{1-\tilde{q}^{kN+n}}\right]
\end{aligned}
\end{equation}
so at the large $N$ limit $\lambda_{n} = -\frac{N}{\ck\tau}\sum_{a=1}^{3}\left[\frac{\tilde{y}_{a}^{n}-\left(\tilde{q}/\tilde{y}_{a}\right)^{n}}{1-\tilde{q}^{n}}+\frac{\tilde{y}_{a}^{N-n}-\left(\tilde{q}/\tilde{y}_{a}\right)^{N-n}}{1-\tilde{q}^{N-n}} + O\left(\tilde{y}_{a}^{N},\left(\tilde{q}/\tilde{y}_{a}\right)^{N}\right)\right]$ where the corrections are exponentially small in $N$. Moreover\footnote{For the first branch we used here the fact that $\dlt_3 = 2\tau - \dlt_1 - \dlt_2 - 1$, the shift property of $\cG$ \eqref{eq:shifting cG} such that we first shift $\dlt_{1,2}$ inside $\cG$ to their bracketed value and shift $\dlt_3 \to 2\tau - [\bdlt_1]_{\ck\tau} - [\bdlt_2]_{\ck\tau} - 1$ accordingly to cancel that, and then shift it again by $\dlt_3$ by additional $2r$ to get to its bracketed value, $[\bdlt_3]_{\ck\tau} = 2\ck\tau - [\bdlt_1]_{\ck\tau} - [\bdlt_1]_{\ck\tau} - 1$. For the second branch, there's one additional shift at the end, resulting in the relative minus sign.}, $b_N = \pm\frac{1}{\ck\tau} + \frac{1}{\ck\tau}\sum_{a=1}^3 \cG\left(0,\frac{N[\dlt_a]}{\ck\tau};-\frac{N}{\ck\tau}\right)$, where the upper sign is for the first case \eqref{eq:def 1st 2nd case} and the lower for the second. Note that all the exponents here are of $\frac{\dlt_{a}}{\ck\tau}$, like the combination that match the D-branes that wrap an $S^{3}\subset S^{5}$, but without the corresponding factor of $N$. We can now evaluate the the determinant and find that there are no perturbative corrections in $1/N$.
\begin{equation}
\begin{aligned}
    \log \det \left[\mathbf{1} + \frac{b_N^{-1}}{N}\cB^{-1}\cC \right]\Bigg|_{m=1} &= \sum_{n=1}^{N-1}\log \left(1 + \frac{b_N^{-1}}{N}\lambda_n \right) \\ 
    &= \sum_{n=1}^{N-1}\log \left[1 \mp \sum_{a=1}^{3}\left(\frac{\tilde{y}_{a}^{n} - \left(\tilde{q}/\tilde{y}_{a}\right)^{n}}{1-\tilde{q}^{n}} + \frac{\tilde{y}_{a}^{N-n} - \left(\tilde{q}/\tilde{y}_{a}\right)^{N-n}}{1-\tilde{q}^{N-n}}\right) \right] \\ &\qquad\qquad + O\left(\tilde{y}_{a}^{N},\left(\tilde{q}/\tilde{y}_{a}\right)^{N}\right) \\
    &= 2\sum_{n=1}^{N-1}\log \left[1 \mp \sum_{a=1}^{3}\frac{\tilde{y}_{a}^{n} - \left(\tilde{q}/\tilde{y}_{a}\right)^{n}}{1-\tilde{q}^{n}} \right] + O\left(\tilde{y}_{a}^{N},\left(\tilde{q}/\tilde{y}_{a}\right)^{N}\right) \\
    &= 2\sum_{n=1}^{\infty}\log \left[1 \mp \sum_{a=1}^{3}\frac{\tilde{y}_{a}^{n} - \left(\tilde{q}/\tilde{y}_{a}\right)^{n}}{1-\tilde{q}^{n}} \right] + O\left(\tilde{y}_{a}^{N},\left(\tilde{q}/\tilde{y}_{a}\right)^{N}\right)\;.
\end{aligned}
\end{equation}
Eventually, one finds
\begin{equation}
\label{eq:expansion determinant at large N m=1}
\boxed{
\begin{aligned}
\det \left[\mathbf{1} + \frac{b_N^{-1}}{N}\cB^{-1}\cC \right]\Bigg|_{m=1} &= \prod_{n=1}^\infty \left[1 \mp \sum_{a=1}^{3}\frac{\tilde{y}_{a}^{n} - \left(\tilde{q}/\tilde{y}_{a}\right)^{n}}{1-\tilde{q}^{n}} \right]^2 \\ 
&\times \left(1 + O\left(\tilde{y}_{a}^{N},\left(\tilde{q}/\tilde{y}_{a}\right)^{N}\right) \right)\;, \quad \text{\first/\second \ case.}
\end{aligned}
}
\end{equation}

\section{Special functions}
\label{app:Special functions}
Throughout we will sometimes use\footnote{Note that this convention is more common in the literature concerning the modular functions and transformations. In some of the literature concerning elliptic functions one uses $q' = e^{\pi i \tau} = \sqrt{q}$, even though it is denoted by $q$ there.}
\begin{equation}
    q = e^{2i\pi\tau} \;, \qquad p = e^{2\pi i \sigma} \;, \qquad z = e^{2\pi i u}
\end{equation}

\paragraph{q-Pochhammer symbol}
The q-Pochhamemer symbol is
\begin{equation}
    (z;q)_n = \prod_{k=0}^{n-1}(1-zq^k) \;,\qquad (z;q)_\infty = \prod_{k=0}^\infty (1-zq^k) \qquad \text{for } |q|<1 \;.
\end{equation}
There are also series expansion and Plethystic representation for $(z;q)_\infty$,
\begin{equation}
    (z;q)_\infty = \sum_{n=0}^\infty \frac{(-1)^n q^{\frac{1}{2}n(n-1)}}{(q;q)_n}z^n = \exp\left[-\sum_{k=1}^\infty \frac{1}{k} \frac{z^k}{1-q^k} \right] \;,
\end{equation}
where the first converges for $|q|<1$ while the second converges for $|z|,|q|<1$.

By relating the symbol to the Dedekind eta function,
\begin{equation}
    \eta(\tau) = e^{\frac{\pi i \tau}{12}}(q;q)_\infty \;,
\end{equation}
one obtains the properties of the q-Pochhammer symbol under modular transformations
\begin{equation}
    (\tilde q; \tilde q)_\infty = \sqrt{-i\tau} e^{\frac{\pi i}{12}(\tau + 1/\tau}(q;q)_\infty \;,
\end{equation}
where $\tilde q = e^{-2\pi i/\tau}$. Finally, we have the asymptotic behaviour
\begin{equation}
    (z;q)_\infty \sim 1-z \qquad \text{for } q \to 0 \;, \qquad \qquad \log(z;q)_\infty \sim -\frac{z}{1-q} \quad \text{for } z\to 0 \;.
\end{equation}

\paragraph{Elliptic theta function}
The elliptic theta function is 
\begin{equation}
\label{eq:elliptic_theta_definition}
    \theta_0(u;\tau) = (z;q)_\infty (q/z;q)_\infty = \prod_{k=0}^\infty (1-zq^k)(1-z^{-1}q^{k+1}) \; ,
\end{equation}
which gives an analytic function on $|q|<1$ with simple zeroes at $z=q^k$ for $k\in \bZ$ and no singularities. The infinite product is convergent in the whole domain. We can also give a plethystic expansion
\begin{equation}
    \theta_0(u;\tau) = \exp\left[ -\sum_{k=1}^\infty \frac{1}{k} \frac{z^k + (qz^{-1})^k}{1-q^k} \right] \;,
\end{equation}
which converges for $|q| < |z| < 1$. The periodicity relations are 
\begin{equation}
\label{eq:elliptic theta periodicity}
\begin{aligned}
    \theta_0(u + n + m\tau; \tau) &= (-1)^m e^{-\pi i m (2  u + (m-1)\tau)}\theta_0(u;\tau) \;, \qquad\qquad m,n\in bZ \\
    \theta_0(u;\tau) &= \theta_0(\tau - u;\tau) = -e^{2\pi i u} \theta_0(-u;\tau) \;,
\end{aligned}
\end{equation}
and under modular transformations
\begin{equation}
    \theta_0(u;\tau + 1) = \theta_0(u;\tau) \;, \qquad \theta_0\left(\frac{u}{\tau}; -\frac{1}{\tau} \right) = -ie^{\frac{\pi i}{\tau}(u^2 + u + \frac{1}{6}) - \pi i u + \frac{\pi i \tau}{6}}\theta_0(u;\tau) \;. 
\end{equation}

One could define the logarithmic derivative of the elliptic theta function 
\begin{equation}
    \label{eq:log_der_elliptic_theta_definition}
    \Theta_0(u;\tau) = \partial_u \log\theta_0(u;\tau) = \frac{\partial_u\theta_0(u;\tau)}{\theta_0(u;\tau)} \;,
\end{equation}
with the series expansion
\begin{equation}
    \label{eq:log_der_elliptic_theta_series_expansion}
    \Theta_0(u;\tau) = 2\pi i\sum_{k=1}^{\infty}\left[\frac{q^{k}z^{-k-1}}{1-q^{k}}-\frac{z^{k-1}}{1-q^{k}}\right] = \frac{2\pi i}{z} \sum_{m=0}^\infty \left[ \frac{z^{-1}q^{m+1}}{1-z^{-1}q^{m+1}} - \frac{zq^m}{1-zq^m} \right] \; .
\end{equation}

Finally, the elliptic theta function can be related to the Jacobi theta functions via
\begin{equation}
\label{eq:ellitpic_theta_jacobi_theta_relation}
    \theta_{1}\left(u;\tau\right) =  ie^{\pi i (\frac{\tau}{4} - u)}\left(q;q\right)_{\infty} \theta_0(u;\tau) \;.
\end{equation}

\paragraph{The elliptic Gamma function}
The elliptic Gamma function is defined by
\begin{equation}
    \wt \Gamma(u;\sigma,\tau) = \prod_{m,n = 0}^\infty \frac{1-p^{m+1}q^{n+1}z^{-1}}{1-p^m q^n z} \; .
\end{equation}
This definition gives a meromorphic single value function on $|p|,|q|<1$ with simple zeroes at $z = p^{m+1}q^{n+1}$ and simple poles at $z = p^{-m}q^{-n}$ for $m,n \ge 0$. The infinite product is convergent on the whole domain. We can also give a plethystic definition
\begin{equation}
    \wt \Gamma(u;\sigma,\tau) = \exp\left[\sum_{k=1}^\infty \frac{1}{k}\frac{z^k - (pqz^{-1})^k}{(1-p^k)(1-q^k)} \right] \;,
\end{equation}
which converges for $|pq|<|z|<1$. The function has the following periodicy relations
\begin{equation}
\label{eq:elliptic gamma quasi periodicity}
\begin{aligned}
    \wt\Gamma(u;\sigma,\tau) &= \wt\Gamma(u;\tau,\sigma) \;, \\
    \wt\Gamma(u;\sigma,\tau) = \wt\Gamma(u+1;\sigma,\tau) &= \wt\Gamma(u;\sigma+1,\tau) = \wt\Gamma(u;\sigma,\tau+1) \;, \\
    \wt\Gamma(u+\sigma; \sigma, \tau) &= \theta_0(u;\tau) \wt\Gamma(u;\sigma,\tau) \;,  \\
    \qquad \wt\Gamma(u+\tau;\sigma,\tau) &= \theta_0(u;\sigma) \wt\Gamma(u;\sigma,\tau) \;.
\end{aligned}
\end{equation}

Moreover,
\begin{equation}
    \wt\Gamma(u;\sigma,\tau) \wt\Gamma(\sigma + \tau - u;\sigma,\tau) = 1 \;.
\end{equation}
The elliptic Gamma function has $SL(3,\bZ)$ modular properties. For $\sigma, \tau, \sigma/\tau, \sigma+\tau \in \bC \setminus \cR$ there is a "modular formula": \cite{Felder:2000}
\begin{equation}
    \wt\Gamma(u;\sigma,\tau) = e^{-\pi i \cQ(u;\sigma,\tau)}\frac{\wt\Gamma\left(\frac{u}{\tau}; \frac{\sigma}{\tau}, -\frac{1}{\tau}\right)}{\wt\Gamma\left(\frac{u-\tau}{\sigma}; -\frac{1}{\sigma}, -\frac{\tau}{\sigma}\right)} = e^{-\pi i \cQ(u;\sigma,\tau}\frac{\wt\Gamma\left(\frac{u}{\sigma}; -\frac{1}{\sigma}, \frac{\tau}{\sigma}\right)}{\wt\Gamma\left(\frac{u-\sigma}{\tau}; -\frac{\sigma}{\tau}, -\frac{1}{\tau}\right)} \;,
\end{equation}
where $\cQ(u;\sigma,\tau)$ is the cubic polynomial
\begin{equation}
    \cQ(u;\sigma,\tau) = \frac{u^3}{3\sigma\tau} -\frac{\sigma+\tau-1}{2\sigma\tau}u^2 +
    \frac{\sigma^2+\tau^2+3\sigma\tau-3\sigma-3\tau+1}{6\sigma\tau}u + \frac{(\sigma+\tau-1)(\sigma+\tau-\sigma\tau)}{12\sigma\tau} \;.
\end{equation}

In the degenerate case $\sigma = \tau$ the formula above is not valid. For $u \in \bC \setminus (\bZ + \tau \bZ)$, however, there is a degenerate relation
\begin{equation}
    \wt\Gamma(u;\tau,\tau) = \frac{e^{-\pi i \cQ(u;\tau,\tau)}}{\theta_0(\frac{u}{\tau};-\frac{1}{\tau})}\prod_{k=0}^\infty\frac{\psi\left(\frac{k+1+u}{\tau}\right)}{\psi\left(\frac{k-u}{\tau}\right)} \;,
\end{equation}
the function $\psi$ is the elliptic digamma function defined below and the polynomial $\cQ$ reduces to 
\begin{equation}
\label{eq:def Q modular}
    \cQ(u;\tau,\tau) = \frac{(2u-2\tau+1)(2u(u+1)-2\tau(2u+1)+\tau^2)}{12\tau^2}\;.
\end{equation}
Using 
\begin{equation}
    \cQ(u+1;\tau,\tau) - \cQ(u;\tau,\tau) = \frac{(u+1)(u+1-2\tau)}{\tau^2}+\frac{5}{6} \;,
\end{equation}
one can check that $\wt\Gamma(u;\tau,\tau)$ is invariant under $u\to u+1$.

\paragraph{Function \matht{\cG}.} We find it convenient to define a function, $\cG$, which is closely related to the logarithmic derivative of $\theta_0$,
\begin{equation}
\cG(u, \Delta;\tau) = \frac1{2\pi i} \, \parfrac{}{u} \log \left(\frac{ \theta_0(\Delta-u; \tau) }{ \theta_0(\Delta + u; \tau)} \right) \;.
\end{equation}
% When not specified, the dependence on $\tau$ is implicit.
It has the series expansion
\be
\label{G series expansion}
\cG(u,\Delta; \tau) = \sum_{\ell=1}^\infty \frac{ (z^\ell + z^{-\ell}) \bigl( y^\ell - (q/y)^\ell \bigr) }{ 1-q^\ell}
\ee
which converges for $|q|< |yz| < 1$ and $|q| < |y/z| < 1$, with $z = e^{2\pi i u}$ and $y = e^{2\pi i \Delta}$. 

The function has the properties
\be
\cG(u, \Delta; \tau) = \cG(-u, \Delta; \tau) = - \cG(u, \tau - \Delta; \tau) \;.
\ee
and also the modular properties, which follow from the ones of $\theta_0$:
\be
\label{G modularity}
\cG(u,\Delta; \tau+1) = \cG(u,\Delta; \tau) \;,\qquad \cG \left( \frac u\tau, \frac\Delta\tau; - \frac1\tau \right) = \tau - 2\Delta -1 + \tau \, \cG(u,\Delta; \tau) \;.
\ee
Its periodicities are
\bea
\label{eq:shifting cG}
\cG(u, \Delta; \tau) &= \cG(u+1, \Delta; \tau) = \cG(u+\tau, \Delta; \tau) = \cG(u, \Delta+1; \tau) \\
\cG(u, \Delta+\tau; \tau) &= 2 + \cG(u, \Delta; \tau) \;.
\eea
In particular, $\cG$ is an elliptic function of $u$, and quasi-elliptic of $\Delta$.

\paragraph{Function \matht{\psi}.} Define, for $\im (t) < 0$, the function
\be
\label{function psi definition}
\psi(t) = \exp \left[ t \log \bigl( 1-e^{-2\pi i t} \bigr) - \frac1{2\pi i} \text{Li}_2 (e^{-2\pi i t}) \right] 
= \exp \left[ - \sum_{\ell=1}^\infty \left( \frac t\ell + \frac1{2\pi i \, \ell^2} \right) e^{-2\pi i t \, \ell} \right] \;.
\ee
The branch of the logarithm is determined by its series expansion $\log(1-z) = - \sum_{\ell=1}^\infty z^\ell/ \ell$, whereas $\text{Li}_2(z) = \sum_{\ell=1}^\infty z^\ell / \ell^2$ is the dilogarithm. One can show that the branch cut discontinuities of the logarithm and the dilogarithm cancel in the definition of $\psi(t)$, therefore the latter extends to a meromorphic function on the whole complex plane. Some useful properties of $\psi(t)$ are:
\be
\psi(t) \, \psi(-t) = e^{-\pi i (t^2 - 1/6)} \;,\qquad\qquad \psi(t+n) = (1-e^{-2\pi i t})^n \, \psi(t) \qquad\text{for}\quad n \in \bZ \;.
\ee
In particular, from (\ref{function psi definition}), $\psi(0) = e^{\pi i / 12}$.

%\printbibliography
\bibliographystyle{JHEP}
\bibliography{BAE}

\end{document}